\documentclass[aps,prl,reprint,superscriptaddress]{revtex4-1}

\newcommand{\EuLaY}{\mbox{Eu$_{1-x}$(La$_{0.254}$Y$_{0.746}$)$_{x}$VO$_3$}}
\newcommand{\LaY}{\mbox{$(\mathrm{La}_{0.254}\mathrm{Y}_{0.746})\rm{VO}_{3}$}}
\newcommand{\order}[2]{\mbox{\textit{#1}-{#2}}}
\newcommand{\Tem}[1]{\mbox{$T_{\rm{#1}}$}}
\newcommand{\kido}[2]{\mbox{$d_{#1}^{\, #2}$}}
\newcommand{\bit}[1]{\ifmmode \text{\boldmath{$#1$}} \else \,{\boldmath $#1$}\, \fi}
\newcommand{\sv}{SV}
\usepackage[dvips]{graphicx}

\begin{document}


\title{Effect of cation size variance on spin and orbital order in \EuLaY}


\author{R.~Fukuta}
\affiliation{Department of Physics, Osaka University, Osaka 560-0043, Japan.}
\author{S.~Miyasaka}
\affiliation{Department of Physics, Osaka University, Osaka 560-0043, Japan.}
\author{K.~Hemmi}
\affiliation{Department of Physics, Osaka University, Osaka 560-0043, Japan.}
\author{S.~Tajima}
\affiliation{Department of Physics, Osaka University, Osaka 560-0043, Japan.}

\author{D.~Kawana}
\affiliation{Condensed Matter Research Center and Photon Factory, Institute of Materials Structure Science, High Energy Accelerator Research Organization, Tsukuba 305-0801, Japan.}
\author{K.~Ikeuchi}
\affiliation{Condensed Matter Research Center and Photon Factory, Institute of Materials Structure Science, High Energy Accelerator Research Organization, Tsukuba 305-0801, Japan.}
\author{Y.~Yamasaki}
\affiliation{Condensed Matter Research Center and Photon Factory, Institute of Materials Structure Science, High Energy Accelerator Research Organization, Tsukuba 305-0801, Japan.}
\author{A.~Nakao}
\affiliation{Condensed Matter Research Center and Photon Factory, Institute of Materials Structure Science, High Energy Accelerator Research Organization, Tsukuba 305-0801, Japan.}
\author{H.~Nakao}
\affiliation{Condensed Matter Research Center and Photon Factory, Institute of Materials Structure Science, High Energy Accelerator Research Organization, Tsukuba 305-0801, Japan.}
\author{Y.~Murakami}
\affiliation{Condensed Matter Research Center and Photon Factory, Institute of Materials Structure Science, High Energy Accelerator Research Organization, Tsukuba 305-0801, Japan.}

\author{K.~Iwasa}
\affiliation{Department of Physics, Tohoku University, Sendai 980-8578, Japan.}

\date{\today}
\begin{abstract}
We have investigated the $R$-ion ($R$ = rare earth or Y) size variance effect on spin/orbital order in \EuLaY.
The size variance disturbs one-dimensional orbital correlation in $C$-type spin/$G$-type orbital ordered states and suppresses this spin/orbital order.
In contrast, it stabilizes the other spin/orbital order.
The results of neutron and resonant X-ray scattering denote that in the other ordered phase, the spin/orbital patterns are $G$-type/$C$-type, respectively.
\end{abstract}
\maketitle
Perovskite-type transition-metal oxides $RM$O$_3$ ($R$\,=\,rare earth elements or Y, $M$\,=\,transition metal ones) and the electron- and hole-doped systems show a lot of attractive physical properties related with $d$ electrons and have been investigated extensively\cite{Imada}.
In $RM$O$_3$, electronic and magnetic states can be changed by controlling the $M$-O-$M$ bond angle, doping carrier, applying magnetic and electric fields, and so on\cite{Imada, Tokura}.
Among these, the $M$-O-$M$ bond angle is systematically dependent on the $R$-site ionic radius\cite{MacLean}.
On the other hand, the structural disorder caused by the size mismatch of the $R$-site cations is also used as the controllable parameters\cite{Tokura, Tomioka, Rodriguez, Attfield, Vanitha}.
We describe this disorder as the ``size variance (SV)", expressed by $\left<r^2_i\right>-\left<r_i\right>^2$ ($r_i$ is the $R$-site ionic radius).
In general, the \sv~suppresses the long-range spin, charge, and orbital order, and consequently the critical temperature ($T$) decreases with increasing the \sv\cite{Tokura, Tomioka}.
In nearly half doped manganites, the \sv~effect induces the suppression of long range charge/orbital order (CO/OO) and ferromagnetic one. 
As a result, the manganites undergo the phase transition from the long range CO/OO insulator or ferromagnetic metal to spin glass insulator\cite{Tomioka}.
In this paper, we report on the findings that the \sv~induces instability of one spin/orbital ordered state but stabilizes the other one in $R$VO$_3$.
This system consequently shows the novel spin/orbital phase transition from one long range order to the other long range order by changing \sv. 
The phase transition between the two types of spin/orbital long range orders in $R$VO$_3$ induced by \sv~is new phenomena, and quite different from those in the manganites, where the phase transition from long range spin/charge/orbital ordered states to short range ordered ones occurs by the increase of \sv.

\begin{figure}[t]
\begin{center}
\hspace{2mm}\includegraphics[width=7.5cm]{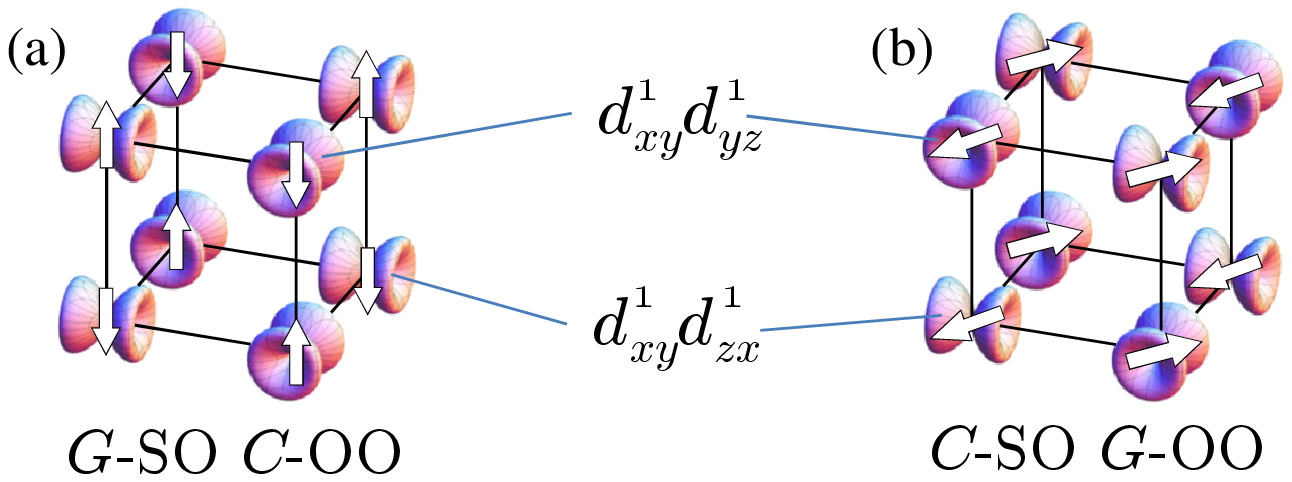}\\
\vspace{1mm}
\includegraphics[width=8.5cm]{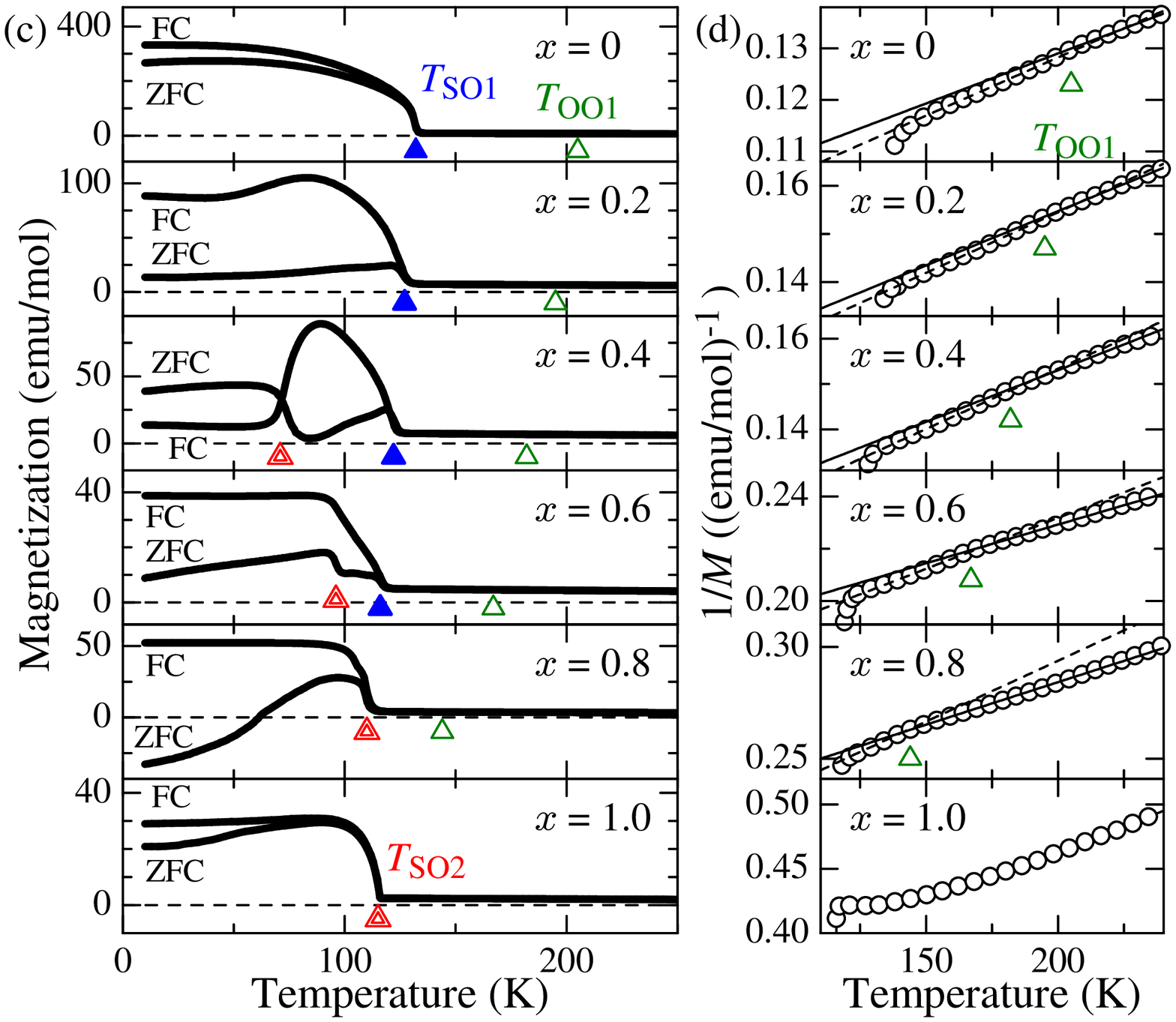}
\caption{
(Color online)
Schematic views of spin (arrows) and orbital order: (a) \order{G}{SO}/\order{C}{OO} and (b) \order{C}{SO}/\order{G}{OO}.
(c) $T$-dependent $M$ for \EuLaY~with applied 1\,kOe in zero-field cooling (ZFC) and field cooling (FC) processes.
The open green, closed blue, and double red triangles indicate \Tem{OO1}, \Tem{SO1}, and \Tem{SO2}, respectively.
(d) $1/M$ as a function of $T$ for these compounds.
Except for $x=1.0$, the dashed lines represent the Curie-Weiss fitting between \Tem{SO1} and \Tem{OO1}, and solid lines above \Tem{OO1}.
}
\label{Magnetization}
\end{center}
\end{figure}

$R$VO$_3$ is one of the prototypical systems with spin and orbital degrees of freedom of $t_{2g}$ electrons.
There are two electrons in the $3d$ orbitals of V$^{3+}$.
One electron always occupies $d_{xy}$ orbital due to the symmetry-lowered crystal field by the orthorhombic distortion, which is coupled ferromagnetically to the other electron in either $d_{yz}$ or $d_{zx}$ orbital through Hund's rule coupling.
$R$VO$_3$ shows two types of the spin/orbital order concomitantly with the structural phase transition, i.e.
$C$-type spin order (\order{C}{SO})/$G$-type orbital order (\order{G}{OO}) with $P2_1/b$ monoclinic structure, and \order{G}{SO}/\order{C}{OO} with $Pbnm$ orthorhombic one\cite{Kawano, Ulrich, Reehuis, Noguchi, Blake, Miyasaka1}.
In \order{C}{OO} the electronic configurations of \kido{xy}{1}\kido{yz}{1}/\kido{xy}{1}\kido{zx}{1} are alternately arranged in $ab$ plane and identical along $c$ axis (Fig. \ref{Magnetization} (a)), while in \order{G}{OO}, those are alternately arranged in all three directions (Fig. \ref{Magnetization} (b)).
$C$-type/$G$-type SO patterns are shown in the same manner as the OO as presented in Figs. \ref{Magnetization} (a) and (b).
The transition $T$ of each SO/OO show systematic dependence on the $r_{i}$\cite{Miyasaka1}.
On the other hand, Yan and co-workers investigated the \sv~effect on the SO/OO order in $R$VO$_3$\cite{Yan}.
They melt-grew Y$_{1-x}$(La$_{0.2337}$Lu$_{0.7663}$)$_{x}$VO$_3$ polycrystals, where the average $r_{i}$ is fixed to be the same as that of Y$^{3+}$.
In this system, the change of the V-O-V bond angle is vanishingly small, while the variance linearly increases with $x$.
With increasing $x$, the transition $T$ of \order{G}{SO}/\order{C}{OO} is enhanced while those of \order{C}{SO}/\order{G}{OO} are suppressed.
Their results have also suggested that in large variance region, there is only one magnetically ordered state.
But there are few information about this state.

In $R$VO$_{3}$, the \sv~of $R$-ions seems not only to destabilize one long-range SO/OO, but also to stabilize the other order.
To clarify the \sv~effect in this $t_{2g}$ orbital system, we have investigated the SO/OO in \EuLaY~single crystal.
The pure material ($x=0$) EuVO$_3$ undergoes one pattern of SO/OO order,
i.e. a structural phase transition from orthorhombic to monoclinic lattices accompanied with \order{G}{OO} at $204\, \mathrm{K}$ and also a magnetic transition from paramagnetic to \order{C}{SO} at $131\, \mathrm{K}$, as $T$ is lowered\cite{Tung, Fujioka1}.
This compound is located near the phase boundary between two SO/OO states.
So we expected that if the large \sv~actually stabilizes the \order{G}{SO}/\order{C}{OO}, the \order{G}{SO}/\order{C}{OO} would appear in \EuLaY.
In this compounds, the average ionic radius of $(\rm{La}_{0.254}\rm{Y}_{0.746})^{3+}$ is the same as that of Eu$^{3+}$, and the variance linearly increases with $x$.
Here, the ionic radii for 12 coordination are $1.228$, $1.360$, and $ 1.183$\,\AA~for Eu$^{3+}$, La$^{3+}$, and Y$^{3+}$, respectively\cite{comment1}.
The choice of this mixed-crystal of EuVO$_3$ and \LaY~enables us to systematically investigate the spin/orbital phase diagram with varying the \sv, while avoiding the large influence of the $4f$-moment on V $3d$ spins because of small magnetic moment for Eu$^{3+}$, and no magnetic La$^{3+}$ and Y$^{3+}$\cite{Fujioka1, Miyasaka2}.
In this study, we have measured the magnetization ($M$), heat capacity ($C$), lattice constants, resonant X-ray scattering (RXS), and neutron diffraction (ND) to clarify the spin/orbital patterns and the phase diagram of $R$VO$_3$ with changing the \sv.

All the samples of \EuLaY~were single crystals grown by a floating-zone method\cite{Miyasaka3}.
The $R$ concentrations of the grown crystals were confirmed to be the same as the nominal ones by ICP analysis.
$M$ measurements were carried out with a SQUID magnetometer.
$C$ measurements were performed by relaxation method.
The lattice constants were measured by the powder X-ray scattering techniques using the X-ray with the energy of 15 keV at BL-8A of Photon Factory (PF), KEK, Japan.
The structural parameters were determined by a Rietveld analysis using the RIETAN-FP\cite{Izumi}.
All the peaks of powder X-ray diffraction profiles for $x=0,\,0.4$ and $1.0$ can be well fitted by using structural parameters of $Pbnm$ orthorhombic or $P2_{1}/b$ monoclinic lattices, and does not show any broadening and splitting. 
The results indicate that this system has no new structural phase and no phase separation. 
The RXS measurements for single crystals of $x=1.0$ with a (1\,0\,0) surface were performed at BL-4C of PF.
In order to analyze whether the polarization of the scattering beam is parallel ($\pi '$ polarization) or perpendicular ($\sigma '$ one) to the scattering plane, we used a pyrolitic graphite (0\,0\,4) crystal.
The ND measurements for the single crystal with $x=1.0$ were performed by using the triple-axis spectrometers of thermal neutron TOPAN (6G), which are installed in the research reactor JRR-3 of Japan Atomic Energy Agency, Tokai.

$T$ dependence of $M$ and its inverse ($1/M$) for \EuLaY~with various $x$ are shown in Figs. \ref{Magnetization} (c) and (d).
As shown in Fig. \ref{Magnetization} (d), the $M$ except for $x=1.0$ follows Curie-Weiss rule in paramagnetic phase.
The slope of $1/M$ changes at the transition $T$ of \order{G}{OO} (\Tem{OO1})\cite{Ren}.
In Fig. \ref{Magnetization} (c), $M$ of EuVO$_3$ ($x=0$) shows a significant jump at $132\, \mathrm{K}$ which we assigned as the transition $T$ of \order{C}{SO} (\Tem{SO1})\cite{Tung, Fujioka1}.
The same magnetic transition has been observed below $x\sim 0.8$.
In addition, a new anomaly in $T$-dependent $M$, which we defined as \Tem{SO2}, appears in the samples with $x\sim 0.4$, and \Tem{SO2} increases with $x$.
At \Tem{SO2}, the antiferromagnetic spin pattern is changed from $C$-type to the other one, as described later.
With increasing $x$, the magnitude of $M$ at low $T$ decreases, which may be related with the decrease of magnetic Eu$^{3+}$.

\begin{figure}[t]
\begin{center}
\includegraphics[width=8cm]{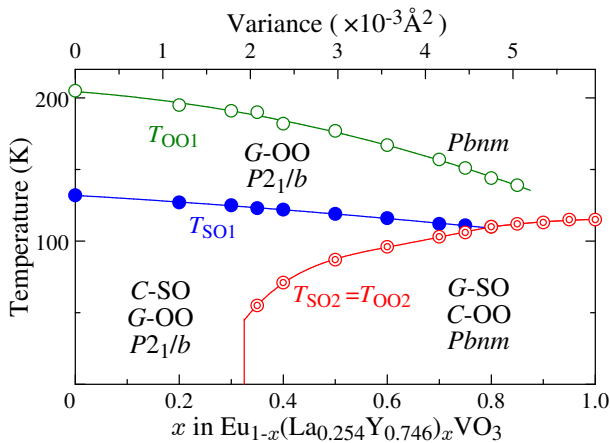}
\caption{
(Color online)
Spin/orbital phase diagram of \EuLaY.
Open green, closed blue, and double red circles indicate the transition $T$ \Tem{OO1}, \Tem{SO1}, and \Tem{SO2} = \Tem{OO2}, respectively.
The top axis shows the variance of the ionic radius on the rare-earth site.
}
\label{Phase}
\end{center}
\end{figure}

Figure \ref{Phase} presents the spin/orbital phase diagram of \EuLaY~determined by the results in Figs. \ref{Magnetization} (c) and (d).
\Tem{OO1} and \Tem{SO1} decrease with increasing $x$ and vanish above $x\sim 0.8$.
As described later, the new anomaly in Fig. \ref{Magnetization} accompanies not only the \order{G}{SO} but also the \order{C}{OO}, and this transition $T$ ($\Tem{SO2}=\Tem{OO2}$) appears above $x\sim 0.4$ and is gradually enhanced with $x$.
This suggests that the \order{G}{SO}/\order{C}{OO} is stabilized by the \sv.
With further increasing $x$, only \order{G}{SO}/\order{C}{OO} exists near $x=1.0$.

\begin{figure}[t]
\begin{center}
\includegraphics[width=8.5cm]{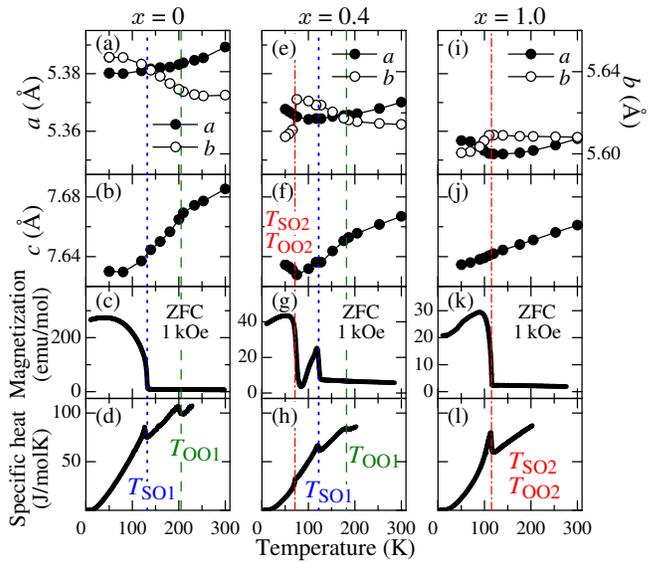}
\caption{
(Color online)
$T$ dependence of the lattice constants \textit{a} and \textit{b} ((a), (e), (i)), \textit{c} ((b), (f), (j)), $M$ measured in 1\,kOe after ZFC ((c), (g), (k)), and $C$ ((d), (h), (l)) for \EuLaY~with $x=0,\,0.4,\,$and $1.0$.
Vertical dashed green, dotted blue, and dash-dotted red lines indicate \Tem{OO1}, \Tem{SO1}, and \Tem{SO2} = \Tem{OO2}, respectively.
}
\label{Lattice}
\end{center}
\end{figure}

Figure \ref{Lattice} shows the $T$ dependence of lattice constants $a,\, b,\, c$ as well as $M$ and $C$ for $x=0, \, 0.4$, and $1.0$ samples\cite{comment2}.
At \Tem{OO1}, $T$ dependence of lattice constants have kinks, and the $C$ has a peak, indicating the structural phase transition from the orthorhombic $Pbnm$ to monoclinic $P2_1/b$ lattices concomitantly with the \order{G}{OO} as decreasing $T$ (Figs. \ref{Lattice} (a), (b), (d)-(f), and (h)).
On the other hand, the anomaly of $M$ and peak of $C$ at \Tem{SO1} indicate the magnetic transition to \order{C}{SO} (Figs. \ref{Lattice} (c), (d), (g), and (h)).
As shown in Figs. \ref{Lattice} (e)-(l), lattice constants, $M$, and $C$ have anomalies at $\Tem{SO2}=\Tem{OO2}$ for $x=0.4$ and $1.0$ samples.
These results reveal that the magnetic and structural phase transitions occur together at \Tem{SO2}.
The changes of lattice constants are very similar to the orbital phase transition to \order{C}{OO}\cite{Bizen}.
These behaviours suggest the existence of the lower-$T$ orthorhombic phase with the \order{C}{OO} (and coupled to \order{G}{SO}).

\begin{figure}[t]
\begin{center}
\includegraphics[width=8.6cm]{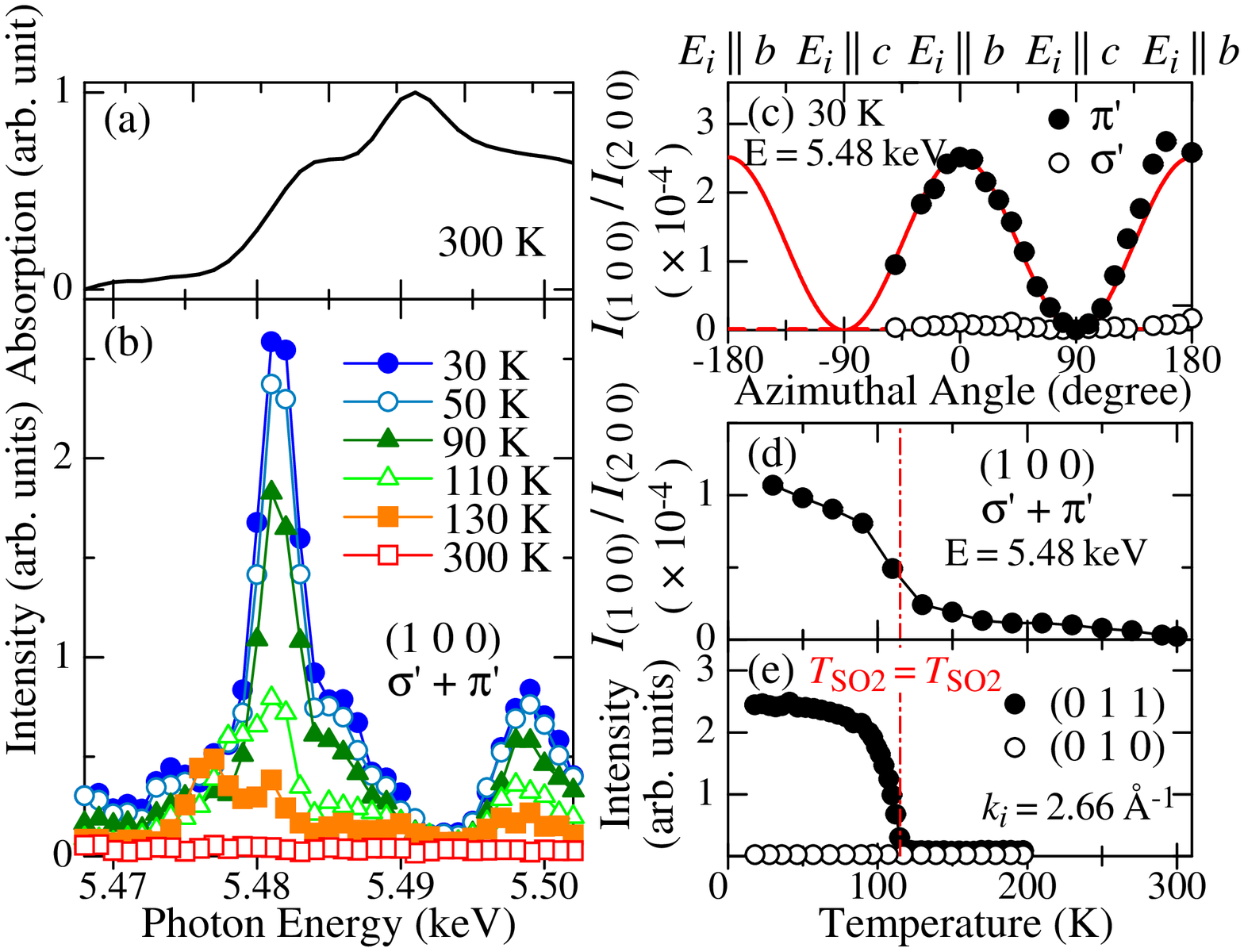}
\caption{
(Color online)
Diffraction results for $x=1.0$ sample of \EuLaY.
(a) X-ray absorption spectrum at 300 K.
(b) Photon-energy spectra of the $\sigma'+\pi'$ component of (1\,0\,0) orbital reflection at various $T$.
The incident beam has $\sigma$ polarization and $E_{i}$ is parallel to the $b$ axis.
(c) Closed and open circles indicate the intensity of the $\pi'$ and $\sigma'$ components of (1\,0\,0) reflection normalized by (2\,0\,0) fundamental one as a function of azimuthal angle.
Red solid and dashed lines are model calculations for \order{C}{OO} of the $\pi'$ and $\sigma'$ components.
(d) $T$ dependence of intensity of the $\sigma'+\pi'$ component of the normalized (1\,0\,0) reflection at 5.48\,keV.
Vertical lines indicate \Tem{SO2} = \Tem{OO2}.
(e) $T$ dependence of intensity of (0\,1\,1) and (0\,1\,0) magnetic Bragg reflections, indicated by closed and open circles, respectively.
$k_{i}=2.66$\AA$^{-1}$ is the momentum of incident neutrons.
}
\label{RXS}
\end{center}
\end{figure}

To confirm the pattern of SO/OO at the ground state above $x=0.35$, we performed the RXS and ND measurements for the $x=1.0$ sample.
Fig. \ref{RXS} (a) shows the X-ray absorption spectrum at 300 K for $x=1.0$.
The rising of the absorption spectrum near 5.48 keV is due to the V \textit{K} main edge.
Fig. \ref{RXS} (b) shows the energy dependence of the intensity of (1\,0\,0) orbital reflection at various $T$.
This reflection does not obey the extinction rule but corresponds to the propagation vectors for \order{C}{OO}.
The symmetry of the degenerated orbital states at each V$^{3+}$ site can be investigated through azimuthal angle dependence of orbital reflections.
As shown in Fig. \ref{RXS} (c), the (1\,0\,0) reflection normalized by the (2\,0\,0) fundamental one at the main edge is maximum with the electric vector of incident beam $E_{i}\parallel b$ configuration and nearly vanishes with $E_{i}\parallel c$.
Model calculations for $\pi '$ and $\sigma '$ components in \order{C}{OO} are also shown in Fig. \ref{RXS} (c).
(In a simple \order{G}{OO} model, $\pi '$ and $\sigma '$ components are equal to zero.)
In our calculations, the atomic scattering tensor of each V$^{3+}$ ion is the same as that in Ref.~\cite{Noguchi} and the structural distortion, estimated by the Rietveld analysis for $x=1.0$, is considered.
The good agreement between the experimental and calculated data of each component, shown in Fig. \ref{RXS} (c), indicates the existence of \order{C}{OO}. \par

Figure \ref{RXS} (d) displays the $T$ dependence of the normalized intensity of the (1\,0\,0) orbital reflection with the $E_{i}\parallel b$ configuration at the main edge in a warming run.
Figure \ref{RXS} (e) demonstrates a comparison between $T$ dependences of the integrated intensity of (0\,1\,1) and (0\,1\,0) magnetic Bragg reflections measured by ND in a warming run.
The (0\,1\,1) reflection is caused by the \order{G}{SO}, and the (0\,1\,0) by the \order{C}{SO}\cite{Kawano}.
The intensity of the (1\,0\,0) orbital reflection and (0\,1\,1) magnetic Bragg reflection are suddenly reduced around $\Tem{SO2} =\Tem{OO2}$ as $T$ is increased.
These results indicate that the \order{G}{SO}/\order{C}{OO} occurs below $\Tem{SO2} =\Tem{OO2}$ in the $x=1.0$ sample.
The intensity of (1\,0\,0) orbital reflection remains finite even above $\Tem{SO2} =\Tem{OO2}$, which is caused by the \order{C}{OO} fluctuation and perhaps the RXS component from local crystal symmetry of space group $Pbnm$.

The \sv~generally suppresses the long range order of spin, orbital, and charge.
The gradual decrease of \Tem{OO1} and \Tem{SO1} with the variance reveals that the \sv~destabilizes \order{C}{SO}/\order{G}{OO} in \EuLaY.
However, the appearance of \order{G}{SO}/\order{C}{OO} and the increase of $\Tem{SO2}=\Tem{OO2}$ with increasing the \sv~can not be explained by the general scenario.
To account for this anomalous behavior of \order{G}{SO}/\order{C}{OO}, one may remember the $R$-site covalency effect\cite{Mizokawa1, Mizokawa2, Bizen}.
In \EuLaY, however, the $R$-site covalency is almost constant, because the average ionic radius of $R$-site is unchanged. 

On the other point of view, the competition of two SO/OO states in $R$VO$_3$ is emphasized\cite{Mizokawa2}.
Among these states, the \order{C}{SO}/\order{G}{OO} includes the quasi-one-dimensional (quasi-1D) orbital chain along $c$-axis, which is described by quasi-1D model for orbital pseudospin ($\tau$=1/2).
This quasi-1D antiferroic orbital chain strongly enhanced the quantum orbital fluctuation, and causes 1D optical response and the orbital excitation\cite{Miyasaka3, Ishihara1, Khaliullin, Motome, Miyasaka4, Fujioka2}.
When the \sv~increases, the 1D orbital chain is easily deranged, and 1D orbital correlation (orbital order and fluctuation) is suppressed.
As a result, the \order{C}{SO}/\order{G}{OO} is destabilized but the other competing SO/OO, \order{G}{SO}/\order{C}{OO}, is relatively stabilized.
Therefore, in the present systems the ground state is changed from \order{C}{SO}/\order{G}{OO} to \order{G}{SO}/\order{C}{OO} and $\Tem{SO2}=\Tem{OO2}$ is gradually enhanced with $x$.
The \order{G}{SO}/\order{C}{OO} phase only exists around $x=1.0$ in this system, while the \sv-free $R$VO$_3$ always has the \order{C}{SO}/\order{G}{OO} at low $T$\cite{Miyasaka1}.
This difference between the present system and $R$VO$_3$ without $R$-ion \sv~indicates that the destruction of 1D orbital chain by \sv~of $R$-ions plays an important role of the disappearance of \order{C}{SO}/\order{G}{OO}.

We have clarified the spin/orbital phase diagram for \EuLaY~with controllable cation \sv~by changing $x$.
By increasing \sv, \order{G}{OO} and \order{C}{SO} tend to be destabilized, while the other ordered phase is stabilized.
We confirmed that this spin/orbital ordered state is the $G$-type/$C$-type by using ND and RXS measurements, respectively.
The \sv~disturbs 1D orbital chain in the state.
The suppression of 1D orbital correlation causes the phase transition from \order{C}{SO}/\order{G}{OO} to \order{G}{SO}/\order{C}{OO} states.

This study has been approved by the PF Program Advisory Committee (No.\,2009S2-008).
The ND measurement was performed under the User program conducted by ISSP, Univ. of Tokyo (No.\,10402)).


\begin{thebibliography}{99}
\bibitem{comment1}In $R$VO$_{3}$, the lattice constants show the linear dependence on the $R$-site ionic radius \cite{Martinez, Sage}. Assuming that Eu is trivalent, the lattice constants for \EuLaY~discussed later satisfies this relationship between lattice constants and $R$-ion size.
(Eu$^{2+}$ has much larger ionic radius than that of Eu$^{3+}$.)
This indicates that the present system has only Eu$^{3+}$.
\bibitem{comment2}At room $T$, the lattice constants slightly and linearly decrease with $x$, although the Eu$^{3+}$ and (La$_{0.254}$Y$_{0.746}$)$^{3+}$ have same ionic radius.
This behavior may be also induced by the change of \sv.
\end{thebibliography}
\end{document}